\documentstyle[amssymb,twocolumn,prb,aps]{revtex}

\begin{document}
\title{Polar and non-polar atomic motions in the relaxor ferroelectric PLZT from
dielectric, anelastic and NMR relaxation}
\author{F. Cordero$^{1}$, M. Corti$^{2}$, F. Craciun$^{1}$, C. Galassi$^{3}$, D.
Piazza$^{3}$ and F. Tabak$^{4}$}
\address{$^{1}$ CNR-ISC, Istituto dei Sistemi Complessi,\\
Area della Ricerca di Roma - Tor Vergata, Via del Fosso del Cavaliere 100,
I-00133 Roma, Italy}
\address{$^{2}$ Dipartimento di Fisica "A. Volta" e unit\`a INFM-CNR, Universit\`a di Pavia,
Via Bassi 6, I-27100 Pavia, Italy}
\address{$^{3}$ CNR-ISTEC, Istituto di Scienza e Tecnologia della Ceramica,
Via Granarolo 64, I-48018 Faenza, Italy}
\address{$^{4}$ Department of Physics Engineering, Hacettepe University,
06532 Beytepe, Ankara, Turkey} \maketitle

\begin{abstract}
Dielectric, anelastic and $^{139}$La NMR relaxation measurements have been
made on the relaxor ferroelectric Pb$_{1-3x/2}$La$_{x}$Zr$_{0.2}$Ti$_{0.8}$O$%
_{3}$ (PLZT) with $x=0.22$. The dielectric susceptibility
exihibits the frequency dispersive maximum due to the freezing of
the polar degrees of freedom around $T\simeq 250$ K. The anelastic
and especially NMR relaxation, besides this maximum, indicate an
intense and broad component at lower temperatures, attributed to
rotational modes of the O octahedra, weakly coupled to the polar
modes. It is discussed why such short range rotational
instabilities, known to occur in the Zr-rich rhombohedral region
of the PLZT phase diagram, might appear also in the Ti-rich
region.
\end{abstract}

\section{INTRODUCTION}

Much effort is being spent during the last years in trying to
understand the so-called relaxor state of disordered
ferroelectrics, and notably the perovskite relaxors, with general
formula ABO$_{3}$. The technological interest arises from the fact
that such materials may be prepared with very high values of
dielectric, piezoelectric, electrostrictive and electro-optical
constants in wide temperature ranges, finding applications like
non-volatile and DRAM memories, capacitors, microelectromechanical
devices, electro-optic modulators, optical switches etc.

Such materials have disorder in the cation sublattices that inhibits the
formation of long range ferroelectric order,\cite{Cro87} and present a
phenomenology similar to that found in spin glasses\cite{VJC90} and other
glassy systems, with freezing of the polar fluctuations, separation of field
cooled from zero field cooled susceptibility curves\cite{KFP99} and non
equilibrium phenomena like aging and memory.\cite{CCW00,115} After the early
explanation of the broadened maximum of the dielectric susceptibility in
terms of a distribution of Curie temperatures, according to the local
composition, other models have been proposed, {\it e.g.} in terms of
quenched random fields due to compositional fluctuations,\cite{WKG92} and
spin-glass state induced by interactions among the polar clusters\cite{VJC90}
or random fields.\cite{VR98} Recently, the spherical random-bond
random-field model\cite{PB99} has been developed, as a useful framework to
understand several properties of the relaxor ferroelectrics.

Regarding the microscopic mechanism of the electric polarization, the
focus has been on perovskites having charge disorder in the B sublattice,
like Pb(Mg$_{1/3}$Nb$_{2/3}$)O$_{3}$ (PMN). In these systems detailed
studies of the atomic displacements have been carried out by neutron\cite
{DKC02,JGR04} and synchrotron x-ray\cite{FRJ00} diffraction, and by x-ray
absorption fine structure,\cite{FPG04} while the relevant phonon
instabilities have been studied by inelastic neutron spectroscopy.\cite
{GWY01,VS02,WSY02} Much recent experimental\cite{CGW98,GCP00} and theoretical%
\cite{GCR02} work is devoted even to the well known ferroelectric system PbZr%
$_{1-y}$Ti$_{y}$O$_{3}$ (PZT) near the boundary at $y\simeq 0.48$ between
the Zr-rich rhombohedral and Ti-rich tetragonal phases.

Another perovskite relaxor system used in several applications is
(Pb/La)(Zr/Ti)O$_{3}$ (PLZT), obtained from PZT by substituting Pb$^{2+}$
with La$^{3+}$. Since in this case the charge disorder is in the A
sublattice, many of the considerations valid for B-disordered relaxors do
not hold for PLZT. The most explored region of the PLZT phase diagram is the
one around 65\% of Zr, while only few studies are devoted to the Ti-rich one,%
\cite{DDV93,DXV94,DXV96} which show particularly high quadratic
electro-optic effect and electrostrictive strain, useful for
quadratic electro-optic modulators, optical phase retarders,
electro-optical shutters and electrostrictive actuators.

In this report we present the results of a study of the atomic
dynamics accompanying the freezing of polarization in La- and
Ti-rich PLZT by measuring its dielectric susceptibility, elastic
compliance and NMR relaxation rate of the $^{139}$La nuclei. These
experiments probe the fluctuations of the electric polarization,
of strain and of the environment of the La nuclei, respectively.

\section{EXPERIMENTAL}

Ceramic (Pb$_{1-3x/2}$La$_{x}$)(Zr$_{0.2}$Ti$_{0.8}$)O$_{3}$ (PLZT $100x$%
/20/80), with the charge compensating vacancies in the Pb sublattice, has
been prepared by using the mixed-oxide method according to the following
processing route. The oxide powders were wet ball milled with zirconia balls
in the stoichiometric amount for 24~h, the suspension was then freeze dried,
sieved to 200~$\mu $m and calcined at 850~$^{\rm o}$C for 4 hours. Samples
of the calcined powders with $x=0$, 2, 12 , 22\% were pressed into
bars and sintered at 1250~${^{\rm o}}$C for 2~h, packed with PbZrO$_3$
+ 5wt\% excess ZrO$_2$ in order to maintain a constant PbO activity at the
sintering temperature.

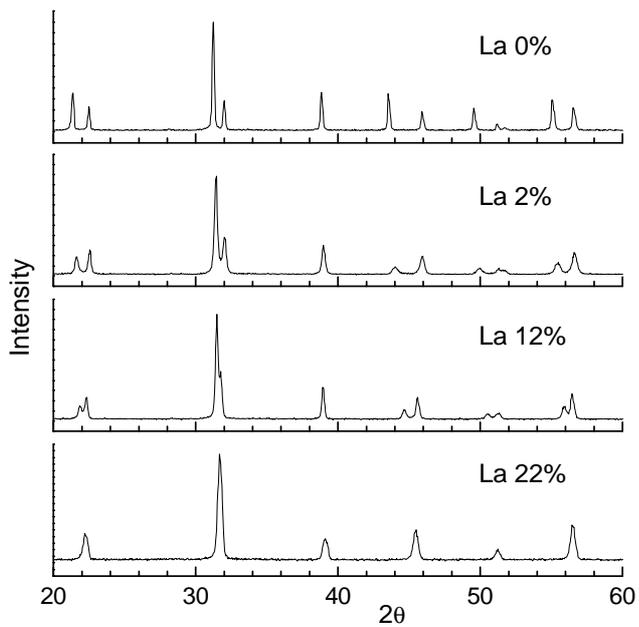
\begin{figure}[htb]
\caption{Diffractograms of PLZT x/20/80 at four La contents.}
\label{fig XRD}
\end{figure}

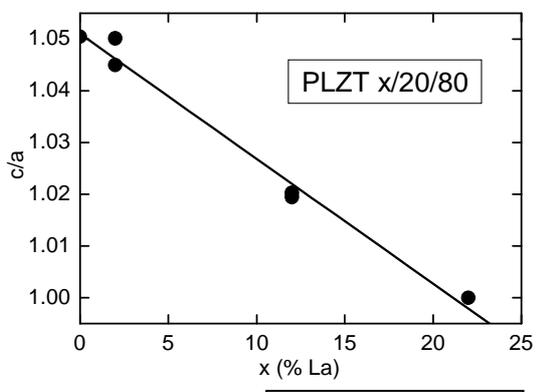
\begin{figure}[htb]
\caption{Vanishing of the tetragonal distortion with increasing La
content.}
\label{fig c/a}
\end{figure}

Phases and microstructure were investigated by x-ray diffraction (XRD)
analysis and by SEM on fracture on a series of samples with $0\le x\le 0.22$.
For $x = 0.22$, the average grain size determined from SEM was 3.2 $\mu $m, and
the density was about 95\% of the theoretical one. The XRD patterns in Fig. \ref
{fig XRD} show the presence of pure perovskitic phase with symmetry varying
from tetragonal to pure cubic as the addition of La increases; the tetragonal
distortion calculated over appropriate peaks of the diffractograms is
shown in Fig. \ref{fig c/a} to vary almost linearly from 5\% for PZT 20/80
to $<0.2$\% for PLZT 22/20/80. The composition under investigation here is PLZT
22/20/80, with a cubic cell and without long range ferroelectric order;
in fact, ferroelectric hysteresis measurements showed a
very slim loop with a remanent polarization of about 1 $\mu $C/cm$^{2}$
(typical values in the ferroelectric state are 30 $\mu $C/cm$^{2}$). Samples
were cut as thin bars approximately $45\times 4\times 0.5$~mm$^{3}$, the
shape required by the anelastic spectroscopy experiments. After cutting, the
samples where annealed for 15~h in air at 750~$^{{\rm o}}$C; the electrodes
for the anelastic and dielectric spectroscopy measurements were applied with
silver paint.

The dielectric susceptibility $\chi =\chi ^{\prime }+i\chi ^{\prime \prime }$
was measured with a HP 4194 A impedance bridge with a four wire probe and a
signal level of 0.5 V/mm, between 200 Hz and 1 MHz. The measurements were
made on cooling at $1-1.5$~K/min from 570 to 200~K.

The complex Young's modulus $E\left( \omega ,T\right) =E^{\prime
}+iE^{\prime \prime }$ was measured on cooling by suspending the bar on thin
thermocouple wires and electrostatically exciting its 1st and 5th flexural
modes, whose resonance frequencies where $\omega _{i}/2\pi \simeq 1.5$ and
20~kHz. The real part of the Young's modulus is related to the resonance
frequency through $\omega _{i}=\alpha _{i}\sqrt{E^{\prime }/\rho }$, where $%
\alpha _{i}$ is a geometrical factor and $\rho $ the mass density. The
elastic energy loss coefficient, or the reciprocal of the mechanical quality
factor,\cite{NB72} is $Q^{-1}\left( \omega ,T\right) =$ $E^{\prime \prime
}/E^{\prime }=$ $s^{\prime \prime }/s^{\prime }$, where $s=s^{\prime
}+is^{\prime \prime }=E^{-1}$ is the elastic compliance; the $Q^{-1}$ was
measured from the decay of the free oscillations or from the width of the
resonance peak. The elastic compliance $s$ is the mechanical analogue of the
dielectric susceptibility $\chi $.

The $^{139}$La NMR spectra and spin-lattice relaxation rates $2W$
have been obtained by standard pulse techniques, operating in the
quadrupole perturbed Zeeman regime, with external magnetic field
$H_{0}$ varying from $1.8$ to $9$ Tesla. The temperature range
explored was $120-600$ K. The spectra were obtained by the Fourier
transform (FT) of the half of the echo signal and also by the
envelope of the echo under sweeping the RF frequency. The echo
signal arises from the powder distribution of the second-order
quadrupole shift of the central line. From FT the central
component showed a width of about $65$ kHz. The satellite
components are spread over a wide frequency range because of the
first-order perturbation effects due to the electric quadrupole
interaction. The pulse length maximizing the $^{139}$La
quadrupolar echo signal was found about a factor $2$ smaller than
the pulse expected for La nuclei in the absence of quadrupole
interaction, indicating that the satellite lines are almost not
irradiated. The relaxation time measurements were performed by
monitoring the recovery of the echo amplitude after irradiation of
the central line with a sequence of saturating pulses in a time
much smaller than the recovery time.  Even though the acquisition
of the recovery of the nuclear magnetization lasted several hours,
aging did not affect the measurements, since the aging effects decrease
with increasing frequency. In order to assess the lack of
influence from aging, the recovery of the nuclear magnetization
was recorded using different sequences of delays and checking that
the amplitude of the signals did not depend on the chosen
sequence.

For the sake of illustration of the relationship between the different
experimental techniques used in the present work, we anticipate here that in
the ideal case that the main contribution to the nuclear relaxation process
originates from the same type of atomic displacements $u\left( t\right) $
that determine the dielectric and elastic uniform susceptibilities, $2W$
could be qualitatively written in the form\cite{Rig84}
\begin{equation}
2W=A^{2}\int \left\langle u\left( t\right) u\left( 0\right) \right\rangle
e^{-i\omega t}dt=A^{2}\,J\left( \omega \right)  \label{2W}
\end{equation}
where $A$ is a constant including the quadrupole moment of the
$^{139}$La nucleus and the derivatives of the electric field
gradient (EFG) components at the nuclear site with respect to the
displacement $u$, while $J\left( \omega \right) $ is the spectral
density of the correlation function for $u(t)$. In this model
situation, a simple relationship between the NMR relaxation rates
and the dielectric data would be possible through the fluctuation-dissipation
theorem, $\chi^{\prime\prime}\left( \omega \right) \propto
\frac {\omega}{T} J\left( \omega \right)$. Later on we shall
discuss the breakdown of such a direct connection.

\section{RESULTS}

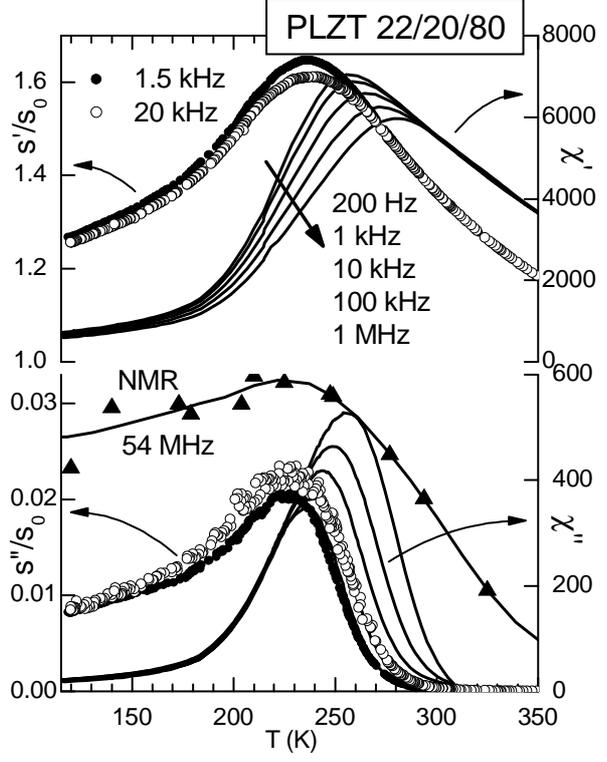
\begin{figure}[htb]
\caption{Real (upper panel) and imaginary parts (lower
panel) of the dielectric susceptibility $\chi$ (right ordinates) and elastic
compliance $s$ (left ordinates), measured at the frequencies indicated in
the figure. Also shown is $\chi _{\text{NMR}}^{\prime \prime }$ defined in
the text measured at 54~MHz.}
\label{fig andielNMR}
\end{figure}

Figure \ref{fig andielNMR} presents the real and imaginary parts of $\chi
\left( \omega ,T\right) $ and $s\left( \omega ,T\right) $, together with one
of the $\chi _{\text{NMR}}^{\prime \prime }\left( \omega ,T\right) $ curves
defined as $\chi _{\text{NMR}}^{\prime \prime }\left( \omega ,T\right)
=\omega \,A^{2}\frac{W}{k_{\text{B}}T}$ according to the aforementioned
simplified model (for $\omega =2\pi \times 54$~MHz). Even though all three
types of relaxations present a maximum in correspondence with the relaxor
transition around 250~K, there are substantial differences in the position
and frequency dispersion of this peak, and in its low temperature tail. The
comparison between dielectric susceptibility and elastic compliance is more
direct, since both involve uniform susceptibilities $\chi \left( q=0,\omega
\right) $ in the same frequency ranges. The differences between the $\chi $
and $s$ curves indicate the presence of additional relaxation modes, besides
the freezing of the polar degrees of freedom probed by $\chi $. For $\chi _{%
\text{NMR}}^{\prime \prime }\left( \omega ,T\right) $ one should take into
account a sizeable contributions from dispersive relaxation modes at
different wave vector which are not included in the uniform susceptibilities.

\subsection{Dielectric susceptibility}

The $\chi ^{\prime }\left( \omega ,T\right) $ curves in Fig. \ref{fig
andielNMR} are in good agreement with the series of $\chi ^{\prime }\left(
\omega ,T\right) $ curves for PLZT $x/20/80$ with $16\le x\le 20$ already
reported,\cite{DXV96} the maximum being slightly smaller and about $\sim 50$%
~K lower in temperature with respect to the highest $x$ reported
there. In order to make a comparison with the NMR $2W$\ curves, we
determine a fitting expression for $\chi \left( \omega ,T\right)
$\ that can be extrapolated at the NMR frequencies. The
interpolation of the dielectric susceptibility involves the choice
of a static susceptibility $\chi \left( 0,T\right) $, and of a
frequency dispersion function, generally taken as a superposition
of elementary relaxation functions with a suitable distribution
$g\left( \ln \tau \right) $ of relaxation times, so that the
dielectric susceptibility is expressed as
\begin{equation}
\chi \left( \omega ,T\right) =\chi \left( 0,T\right) \int d\ln \tau
\,\,g\left( \ln \tau \right) \frac{1}{1-i\omega \tau }\,.  \label{chi}
\end{equation}
In relaxor ferroelectrics, $\chi \left( 0,T\right) $ follows a
modified Curie-Weiss law at temperatures sufficiently far from the
relaxor peak, but there is no divergence on approaching it. For
the present purposes, we limit ourselves to finding a
phenomenological expression for $\chi \left( 0,T\right) $ that is
suitable to fit the data down to temperatures as close as possible
to the relaxor peak. An excellent fit can be obtained with a
logistic function, which saturates at low $T$ instead of
diverging,
\begin{equation}
\chi \left( 0,T\right) =\chi _{0}+\frac{\Delta \chi }{1+\left(
T/T_{0}\right) ^{p}}  \label{chi0}
\end{equation}
with $\chi _{0}=270$, $\Delta \chi =9260$, $T_{0}=319$~K, $p=5.65$. The
value of $\chi _{0}$ seems too high for representing the infinite frequency
limit $\chi _{\infty }$, and is therefore considered as a fitting parameter
in (\ref{chi0}), while $\chi _{\infty }$ is neglected; the fitting curve is
shown in Fig. \ref{fig chi(0,T)}.

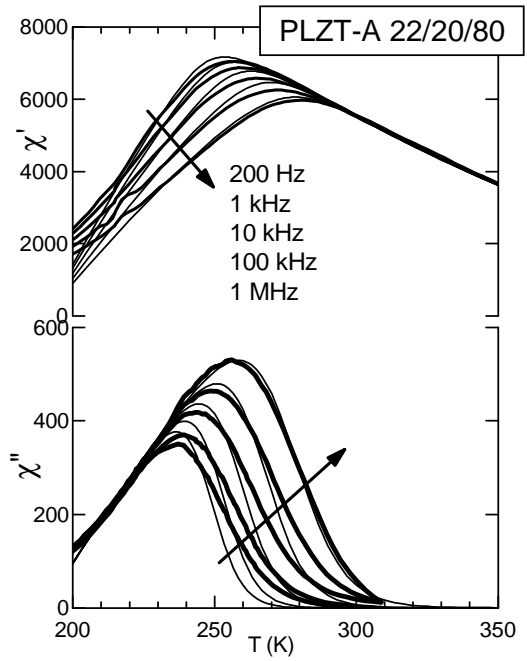
\begin{figure}[htb]
\caption{Real and imaginary parts of the experimental
dielectric susceptibility (thick lines) and fitting curves (thin lines).}
\label{fig fitchi}
\end{figure}

Regarding the frequency dispersion, we tried with
exponential\cite{LC95} and uniform\cite{MRM76,DY90,KBP00}
distribution functions in $g\left( \ln \tau \right) $. The first
type of distribution function was adopted to interpret dielectric
measurements on Pb(Mg$_{1/3}$Nb$_{2/3}$)O$_{3}$,\cite{LC95} but
could hardly reproduce the present data. Instead, we could
interpolate reasonably well both $\chi ^{\prime }$ and $\chi
^{\prime \prime }$ with a modified uniform distribution between
$\ln \tau _{1}$ and $\ln \tau _{2}\left( T\right) $, with the
cutoff at the longest time broadened over $w\left( T\right) $ in
the logarithmic scale:
\begin{eqnarray}
g\left( \ln \tau \right) &=&\left\{
\begin{array}{l}
g_{0}\frac{1}{2}\left[ 1-\tanh \left( \frac{\ln \tau -\ln \tau _{2}}{w}%
\right) \right] \quad {\rm for\,}\tau >\tau _{1} \\
0\quad {\rm for\,}\tau <\tau _{1}
\end{array}
\right.  \label{g} \\
g_{0} &=&2/\,\left[ \ln \tau _{2}/\tau _{1}+w\ln \left[ 2\cosh \left( \frac{%
\ln \tau _{2}/\tau _{1}}{w}\right) \right] \right] \,  \nonumber
\end{eqnarray}
with $w\left( T\right) =0.15\times \left( \ln \tau _{2}\left( T\right) -\ln
\tau _{1}\right) $ scaling with the distribution width. The longest
relaxation time at half maximum of the distribution has been assumed to
follow the Vogel-Fulcher law, similarly to what observed for other
compositions of PLZT\cite{KFP99,KBP00}
\begin{equation}
\ln \tau _{2}=\ln \tau _{0}+E/\left( T-T_{\text{VF}}\right)
\end{equation}
with $\tau _{0}=10^{-14}$~s, $E=1450$~K and $T_{\text{VF}}=190$~K, while the
fastest relaxation time has been set to $\tau _{1}=5\times 10^{-14}$~s. The $%
\chi \left( \omega ,T\right) $ curves have been obtaining by numerically
integrating eq. (\ref{chi}) and are shown in Fig. \ref{fig fitchi}. The fit
is optimized on $\chi ^{\prime \prime }\left( \omega ,T\right) $ at the
highest frequency, in order to obtain a reliable extrapolation of the
spectral density
\begin{equation}
J\left( \omega ,T\right) \propto \frac{T}{\omega }\chi ^{\prime \prime
}\left( \omega ,T\right)
\end{equation}
\ at the NMR frequencies. For $w=0$, $\chi ^{\prime \prime }$ has the
analytical form $\chi ^{\prime \prime }=\chi \left( 0,T\right) \left( \ln
\tau _{2}/\tau _{1}\right) ^{-1}$ $\left[ \arctan \left( \omega \tau
_{2}\right) -\arctan \left( \omega \tau _{1}\right) \right] $, which is
valid also for $w\neq 0$ and $T$ below the maximum, where the shape of the
cutoff at $\omega \tau _{2}\gg 1$ has no influence. The low $T$
approximation is therefore $\chi ^{\prime \prime }\left( \omega ,T\right)
\simeq \chi \left( 0,T\right) $ $\frac{\pi }{2}\left( \frac{T-T_{\text{VF}}}{%
E_{2}-E_{1}}\right) $, where the linear increase with $T$ above $T_{\text{VF}%
}$ comes from the $T$ dependent normalization factor $\left( \ln \tau
_{2}/\tau _{1}\right) ^{-1}$; therefore, considering the weak temperature
dependence of $\chi \left( 0,T\right) $ below the maximum, $T_{\text{VF}}$
is the temperature at which the linear extrapolation of $\chi ^{\prime
\prime }$ below the maximum vanishes. Of course, there is no real divergence
of the distribution width and the dielectric susceptibility does not vanish
below $T_{\text{VF}}$, but the Vogel-Fulcher hypothesis allows the
dispersive maximum to be well described.

\subsection{Elastic compliance}

The elastic compliance curves $s\left( \omega ,T\right) $ measured at the
1st and 5th resonance frequencies (1.5 and 20~kHz) are shown in Fig. \ref
{fig andielNMR} together with the susceptibility curves. The curves are
normalized to the extrapolated value of $s^{\prime }$ at infinite
temperature, $s_{\infty }$, which should be the equivalent of $\chi _{\infty
}$. An important difference between the dielectric and the elastic
susceptibilities, however, is that in relaxor ferroelectrics $\chi _{\infty
} $ brings a negligible contribution to the dielectric response; in fact, $%
\chi _{\infty }$ is expected\cite{GCW99} to be $\lesssim 10$ (even
though $\chi _{\infty }\sim 200$ has ben reported in PLZT 9/65/35\cite
{BKP00}) and here it is $\chi ^{\prime }\sim 7000$ at the maximum. Instead,
the background lattice contribution to $s^{\prime }$ is much more important:
the peak value of $s^{\prime }$ is only 1.6 times larger than $s_{\infty }$.
Since the lattice contribution to $s^{\prime }$ is also temperature
dependent, it is easier to compare only the imaginary parts of the different
susceptibilities, to which $s_{\infty }$ and $\chi _{\infty }$ do not
contribute. The elastic $s\left( \omega ,T\right) $ differs from $\chi
\left( \omega ,T\right) $ in three ways: {\it i)}\ the maxima of the real
and imaginary parts are at lower temperature; {\it ii)} the frequency
dispersion is smaller both in temperature shift and change of intensity, and
the amplitude dependence of $s^{\prime }$ is reversed with respect to $\chi
^{\prime }$; {\it iii)} the low temperature tail is much higher than in the
dielectric case ($s^{\prime \prime }\left( 150~\text{K}\right) /s^{\prime
\prime }\left( T_{\text{max}}\right) =0.5$ for the elastic case at 20 kHz,
while $\chi ^{\prime \prime }\left( 150~\text{K}\right) /\chi ^{\prime
\prime }\left( T_{\text{max}}\right) =0.07$ at 10~kHz). The comparison
between $\chi ^{\prime \prime }$ and $s^{\prime \prime }$ is rather
straightforward, since they both represent uniform susceptibilities. The
only differences are in the relaxation strengths, which are sensitive to
changes in electric dipolar degrees of freedom in one case and elastic
quadrupolar in the other. In the simplest case of independent relaxing
units, e.g. off-centre atoms hopping between positions $i$ and $j$ with a
correlation time $\tau $ one has\cite{NB72}
\begin{equation}
\Delta \chi \left( 0,T\right) =\frac{\left( \Delta p\right) ^{2}}{k_{\text{B}%
}T}\,,\quad \Delta s\left( 0,T\right) =\frac{\left( \Delta \lambda \right)
^{2}}{k_{\text{B}}T}  \label{relstrength}
\end{equation}
where $\Delta {\bf p}={\bf p}^{(i)}-{\bf p}^{(j)}$ is the change of the
electric dipole during the jump, and $\Delta \lambda =\lambda ^{(i)}-\lambda
^{(j)}$ is the change of the elastic quadrupole (generally called elastic
dipole\cite{NB72} for analogy with the electric and magnetic cases). The
relaxation of a same unit may produce a change in ${\bf p}$ but not in $%
\lambda $, e.g. the reorientation of an off-centre unit by 180$^{{\rm o}}$
changes ${\bf p}$ to $-{\bf p}$ but does not affect $\lambda $, which is a
centrosymmetric strain (representable as a strain ellipsoid). Conversely,
there may be relaxation modes affecting $\lambda $ but not ${\bf p}$, e.g.
rotations or distortions of the octahedra. The differences in the dielectric
and elastic susceptibilities indicate that the latter is little affected by
the dipolar relaxation, since the peak in $s^{\prime \prime }$ is not
proportional to the relaxor peak in $\chi ^{\prime \prime }$; in addition,
there are other relaxation modes almost invisible to $\chi $, but
contributing to $s$ below the relaxor transition. These modes are faster and
only loosely follow the dipolar freezing, as indicated by the lower
temperature of the peak and the huge low temperature tail.

The role of electrostrictive strain in driving the anelastic relaxation in
relaxor ferroelectrics has been emphasized,\cite{DXV96,VJC91c} but cannot
account for the present results; in fact, if the strain fluctuations were
simply driven by the dipolar ones through electrostriction, $s^{\prime
\prime }$ would follow $\chi ^{\prime \prime }$ much more closely, since the
electrostrictive coupling is expected to be weakly dependent on temperature.

\subsection{NMR}

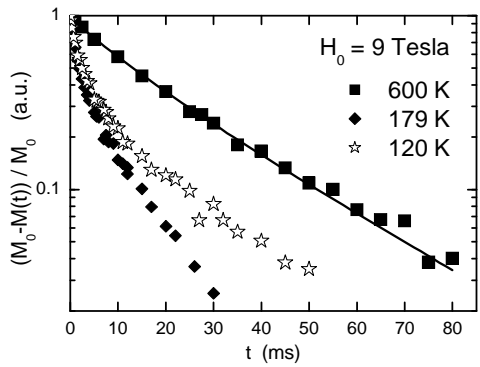
\begin{figure}[tbh]
\caption{Recovery behavior of the NMR echo signal
after a RF pulse sequence saturating the central ($+1/2
\Leftrightarrow -1/2$) transition for different temperatures
(squares 600~K, diamonds 179~K and stars 120~K). At high
temperature the recovery is well described by Eq. (\ref{y}) (solid
line). In the low temperature range the recovery shows the change
over to stretched exponentials.} \label{fig decay}
\end{figure}

Under the conditions of measurements outlined in Section II, by
solving the master equations for $I=7/2$, the recovery of the echo
signal is expected to be described by the law
\begin{eqnarray}
y(t) &=&0.012\,e^{-0.47\,W\,t}+0.714\,e^{-1.33\,W\,t}+  \label{y} \\
&&+0.068\,e^{-2.4\,W\,t}+0.206\,e^{-3.8\,W\,t}  \nonumber
\end{eqnarray}
In deriving Eq. (\ref{y}) the relaxation transition probability $W_{1}$ and $%
W_{2}$ ($\Delta m=1$ and $\Delta m=2$ transitions respectively)
have been assumed practically equal, as usual in powdered
samples.\cite{Rig84} At high temperatures, well above the
ferroelectric relaxor transition
temperature, the recovery of the echo signal is rather well fitted by Eq. (%
\ref{y}), as shown by the solid line in Fig. \ref{fig decay}. One notes
that, according to Eq. (\ref{y}), the relaxation rate $2W$ can also be
obtained from the time $t^{*}$ where $y(t^{*})=0.41$, yielding $%
2W=(t^{*})^{-1}$.

In the low temperature range Eq. (\ref{y}) does no longer describe
the recovery of the echo (see Fig. \ref{fig decay}). This is the
obvious consequence of the fact that, on approaching the relaxor
transition temperature, each exponential in Eq. (\ref{y}) becomes
stretched. A meanigful fitting according to Eq. (\ref{y}) is
hardly possible, and thus in the low temperature range ($T<300$ K)
an effective $2W$ has been extracted as the inverse of the time
$t^{*}$.

The $^{139}$La relaxation rates $2W$ are reported in Fig. \ref{fig
NMRvsT} for three representative measuring frequencies $\omega /2
\pi$. From 700~K, $2W$ increases on decreasing temperature passing
through a maximum and then decreasing again. The maxima in $2W$
occur in the temperature range where the maxima in imaginary part
of the dielectric constant are also detected (see Fig. \ref{fig
andielNMR}). This suggests that, at least in the high temperature
range, the mechanisms responsible of the reorientations of the
electric dipole moments are also involved in the $^{139}$La
relaxation
process. It should be remarked that, by varying the magnetic field from $%
H_{0}=1.8$~T to $H_{0}=9$~T, the maxima of $2W$ do not sizably
shift in temperature. This direct qualitative observation by
itself suggests that the freezing process involves a large
distribution of correlation times, as can be expected for a
strongly disordered relaxor.

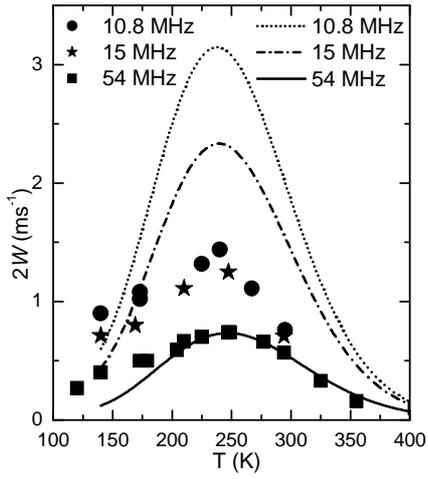
\begin{figure}[tbh]
\caption{Temperature dependence of the NMR relaxation
rates $2W$ for different measuring frequencies. The lines
represent the theoretical curves assuming the LC distribution (see
text).} \label{fig NMRvsT}
\end{figure}

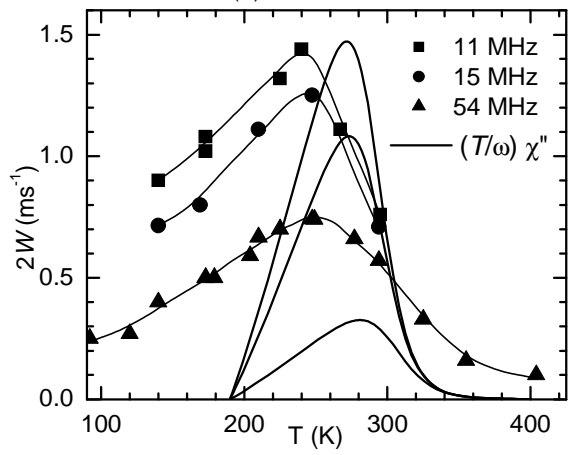
\begin{figure}[tbh]
\caption{Comparison between the NMR relaxation rates $2W$
and the $J(\omega,T)$ curves extrapolated from the fits to
$\chi^{\prime\prime}(\omega,T)$ of Fig. \ref{fig fitchi}.}\label{fig NMRchi}
\end{figure}

In the weak collision approach, for overdamped phonon modes and/or
localized modes in disordered systems (direct relaxation process)
the relaxation rate driven by the time dependence of the EFG
components can be written in the form anticipated in Eq.
(\ref{2W}), where the spectral density $J_{1}(\omega )$ and
$J_{2}(2\omega )$ have been assumed practically coincident. We can
explicitate the constant $A^{2}$ and write $2W$ in the form
$2W\simeq 5\langle \omega _{Q}^{2}\rangle _{fl}J(\omega )$,
$\langle \omega _{Q}^{2}\rangle _{fl}$ being an average effective
strength of the time dependent quadrupole coupling and $J(\omega
)$ the spectral density of the reduced correlation
function.\cite{Rig84} For monodispersive processes, with a single
correlation time, the spectral density would be $J(\omega )=2\tau
/(1+\omega ^{2}\tau ^{2})$. In the fast motions regime, i.e. $\tau
^{-1}\gg \omega $, one would have a frequency independent $2W$ of
the order of $2W\simeq 12\langle \omega _{Q}^{2}\rangle _{fl}\tau
$, while for slow motions $(\tau ^{-1}\ll \omega )$ it is
$2W\simeq 10\,\langle \omega _{Q}^{2}\rangle _{fl}(\omega^{2} \tau
)^{-1}$. From the value of $2W$ at the maximum an order of
magnitude of the fluctuating quadrupole coupling constant can be derived: $%
\langle \omega _{Q}^{2}\rangle _{fl}^{1/2}\simeq $ $(2W_{m}/\omega
)^{1/2}\simeq $ $50-100$~kHz, a reasonable value for the motions
around the equilibrium position of the ions. In the present case,
the $\omega $ dependence of $2W$ is reduced by the broad
distribution of correlation times typical of relaxors. We first
tried with the same uniform distribution function adopted for the
dielectric case, but it is apparent from Fig. \ref {fig NMRchi}
that the extrapolation of the $\left( T/\omega \right) \chi
^{\prime \prime }\left( \omega ,T\right) $ curves to the NMR
frequencies does not  properly reproduce $2W$. We therefore tried
with the distribution proposed by Lu and Calvarin\cite{LC95} (LC)
for their dielectric relaxation data in
Pb(Mg$_{1/3}$Nb$_{2/3}$)O$_{3}$ and Pb$_{2}$KTa$_{5}$O$_{15}$. The
LC distribution is assumed in the form

\begin{equation}
P\left( E\right) =\left\{
\begin{array}{l}
w^{-1}e^{\left( E_{C}-E\right) /w}\quad {\rm for\,}E \ge E_{C} \\
0\quad {\rm for\,}\ E < E_{C}
\end{array}
\right.
\end{equation}
where $E=KV$ is the activation energy for thermal fluctuations of a polar
cluster in the paraelectric matrix, proportional to its volume. $%
E_{C}=KV_{C} $ is the minimum activation energy corresponding to a
critical volume of the order of the volume of a crystalline cell.
The temperature dependence of the distribution width $w$ has been assumed
of the form $w = E_0 e^{E_1/T}$. The
correlation times are related to the activation energies by $\tau
=\tau _{0}e^{E/T}$. In Fig. \ref {fig NMRvsT} the lines represent
the temperature dependence of $2W$
numerically obtained in the assumption of LC distribution with $%
\tau _{0} = 10^{-13}$ s, $E_{C}= 100$ K,  $E_{0}=115$ K and $E_{1}=750$ K.
The fit of the experimental results, although better than the simple
extrapolation of the dielectric ones, is again poor.

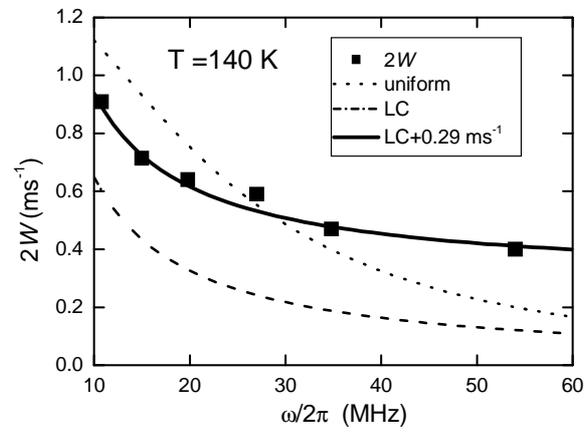
\begin{figure}[htb]
\caption{$2W$ relaxation rates at $T=140$~K as a
function of the measuring frequency (filled squares). The lines
give the frequency dependence expected for the uniform
distribution of the correlation time (dashed line) and for the LC
model (dotted-dashed line). The solid line is obtained by adding
to the LC behavior a frequency independent contribution of 0.29
~ms$^{-1}$.} \label{fig NMRvsf}
\end{figure}

A detailed series of measurements as a function of $\omega $ has
been
carried out at constant temperature, $T=140$ K, below the maxima in $2W$ vs $%
T$. In Fig. \ref{fig NMRvsf} the experimental data at $T=140$ K
have been compared with the frequency dependence expected on the
basis of the relaxation rates according to the LC distribution
(dashed line) and to a uniform distribution (dotted
line). It is evident that both the absolute values as well as the
frequency dependence do not agree with the theoretical
expectations according to the above models. Here we note that a
good fit could be obtained if a frequency independent
contribution, arising from a second type of motion effective in
the relaxation process, is added to the frequency behavior of the
LC distribution (solid line). As noted above, an almost frequency
independent contribution implies a prevalence of fast motions,
with $\omega \tau \ll 1$ still at 140~K, consistent with the broad
contribution at low temperature both in $2W$ and in the anelastic
$s^{\prime \prime }$. We conclude that the $^{139}$La NMR
relaxation contains additional contributions with respect to the
dipolar freezing dominating the dielectric susceptibility,
especially in the low temperature region.

\section{DISCUSSION}

\subsection{Non-polar relaxation modes}

The comparison between dielectric, anelastic and NMR relaxations indicates
the existence of at least two distinct types of atomic motions: {\it i}) polar
modes that dominate the behavior of $\chi ^{\prime \prime }$, giving rise to
the typical relaxor maximum, but are less evident in $s^{\prime \prime }$\
and $2W$; {\it ii}) other modes, affecting much less $\chi $\ and dominating
the nuclear magnetic and anelastic relaxations. The second type of modes
prevails at lower temperature and should be essentially non-polar, since it
does not appear in the dielectric relaxation.

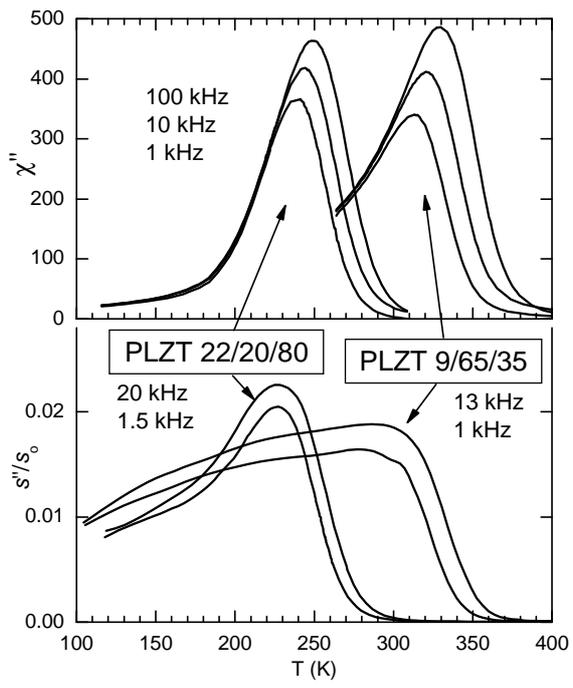
\begin{figure}[htb]
\caption{Comparison between the imaginary parts of the
dielectric susceptibilities (upper panel) and elastic compliances (lower
panel) of PLZT 9/65/35 and PLZT 22/20/80.}
\label{fig tetrarhomb}
\end{figure}

A comparison between dielectric and anelastic relaxation in PLZT 9/65/35,
shows even larger differences,\cite{114} since $s^{\prime \prime }$
completely lacks the relaxor peak and only presents a step-like increase at
a temperature $\sim 30$~K lower than that of the relaxor peak in $\chi
^{\prime \prime }$; this is shown in Fig. \ref{fig tetrarhomb}. It was
proposed that the main difference between the two types of responses
consists in the lack of sensitivity of the elastic compliance to the 180$^{%
{\rm o}}$ switching of the polarization, which would be the main responsible
for the relaxor peak; the non-180$^{{\rm o}}$ polarization dynamics,
contributing also to $s^{\prime \prime }$, would instead have an increasing
role at lower temperature.\cite{114} The NMR data, however, suggest a
different picture; in fact, there is no reason to suppose that the $^{139}$%
La nuclear relaxation should be much more sensitive to
non-180$^{{\rm o}}$ rather than to 180$^{{\rm o}}$ polarization
dynamics, and therefore the explanation of the predominant weight
of relaxation in $2W$ at temperature lower than that of the
relaxor peak can hardly rely on non-180$^{{\rm o}}$ polarization
dynamics. Rather, this indicates that the NMR relaxation probes a
non-polar dynamics, not appearing in $\chi $, presumably the same
type of motion appearing in the anelastic response. Motions of
this type may be rotations or distortions of the octahedra and
antiferrodistortive displacements of the cations.

\subsection{Rotations of the octahedra}

Rotations or deformations of the octahedra would certainly provide a strong
contribution to $2W$, since a La ion is surrounded by 12 nearest neighbor O
atoms at $a/\sqrt{2}$, and 8 second neighbor Ti/Zr atoms at $\sqrt{3}/2\,a$;
the contribution of the atomic shifts to the EFG fluctuations decreases with
the 6th power of the distance from the La site, so that the contribution
from motions of the O octahedra (or of La with respect to the surrounding
octahedra) is enhanced by 5 times over the contribution from the motion of
the Zr/Ti atoms. Since the microscopic polarization in PLZT is mainly
attributed to the off-centre displacements of the cations with respect to
the O octahedra, and there is little correlation between these polar
fluctuations and $2W$, it seems reasonable to assign the main contribution
to the NMR relaxation to rotations or deformations of the octahedra.

Distortions of the octahedra of the Jahn-Teller type are found in
perovskites with cations supporting mixed valence, like manganites, but this
is not the case of ferroelectric PLZT, where both Zr and Ti have fixed
valence $+4$. Jahn-Teller polarons in ferroelectric perovskites may be
observed only after suitable doping,\cite{LSH02} e.g. in Nb-doped BaTiO$_{3}$
where each Nb$^{5+}$ is compensated by one Jahn-Teller active Ti$^{3+}$; our
samples might well contain some minority charged defects, like O vacancies,
but they are insulators and a mechanism like a Jahn-Teller polaron cannot be
considered as source of the intense relaxation observed below the relaxor
transition.

Rotations of the octahedra are instead typical of all perovskites, including
the ferroelectrc ones, and the low temperature phases of Zr-rich PZT are the
result of the condensation of unstable rotational modes of the O octahedra,
together with ferro- and antiferroelectric modes of the cation sublattices.%
\cite{FS01,Leu03} The PZT phase diagram\cite{CGW98,DXV95} grossly consists
of a cubic high temperature phase, which transforms into ferroelectric
tetragonal for a Ti fraction $y>0.48$ and ferroelectric rhombohedral below
the almost temperature independent boundary $y\simeq 0.48$. In the
tetragonal phase, the A (Pb) and B (Zr/Ti) sublattices are shifted along the
$c$ axis with respect to the O octahedra, which are additionally elongated.
This is seen as the result of the condensation of the unstable $\Gamma _{15}$
phonon. In the rhombohedral phase the cations are shifted along the
pseudocubic $\left[ 111\right] $ direction, and, below a certain
temperature, the octahedra rotate about the polar $\left[ 111\right] $ axis,
giving rise to the low-temperature rhombohedral phase. Additional rotational
instabilities of local character have been proposed to exist in Zr-rich PZT%
\cite{Vie95} and especially in PLZT.\cite{DXV95,VXP93} Viehland and
coworkers found $\frac{1}{2}\left\langle 111\right\rangle $ and $\frac{1}{2}%
\left\langle 110\right\rangle $ superlattice spots in electron diffraction
by TEM, which they associate with coordinated rotations of the octahedra by
an angle $\pm \phi $ about the pseudocubic $c$ axis, with successive layers
in phase (condensation of the $M_{3}$ phonon mode\cite{FHY74} producing $%
\frac{1}{2}\left\langle 110\right\rangle $ reflections) or antiphase ($%
R_{25} $ mode producing $\frac{1}{2}\left\langle 111\right\rangle $
reflections). These reflections progressively disappear with increasing Ti
content, and in PLZT 12/40/60 only weak spots were found in a minority
fraction of the selected areas.\cite{VXP93} It has later been argued\cite
{CGW98} that such superlattice reflections might mainly arise from
additional shifts of the Pb atoms along pseudocubic $\left[ 100\right] $
directions, coupled with the octahedral rotations; in addition, since these
reflections are not observed by x-ray or neutron diffraction, the regions of
additional $M_{3}$ or $R_{25} $ ordered rotations might exist only near the
surface of grains, at least in La-free PZT. Rotational instabilities of $%
M_{3}$ and $R_{25}$ type have also been found in first-principle
calculations of the phonon instabilities in rhombohedral PZT.\cite{Leu03}

While the existence of these rotational instabilities is well documented in
the rhombohedral region of the phase diagram of PZT and PLZT, we are not
aware of any such observation in Ti-rich PLZT. Therefore, octahedral
rotations seem at first unlikely to occur in our PLZT 22/20/80, considering
that the parent PZT 20/80 is well within the tetragonal region of the
phase diagram.
We argue, however, that the substitution of Pb with La changes the situation
considerably with respect to PZT. First of all, XRD measurements on a series
of samples with increasing La content (Fig. \ref{fig XRD}) show that the
addition of La suppresses the tetragonal distortion, which is 5\% in PZT
2/20/80, 2\% in PLZT 12/20/80, and $<0.2\%$ in PLZT 22/20/80, implying the
absence of a strong stabilization of the tetragonal phase in the present case.
A similar
behavior has been found in PLZT $x/$40/60.\cite{DXV94} The La addition
should also favor the rotational instability at the expenses of the
tetragonal one. In fact, calculations of the phonon dispersions\cite
{GCW99,FS01} show that both the $\Gamma _{15}$ tetragonal and the rotational
$R_{25}$ instabilities exist in PbTiO$_{3}$ and PbZrO$_{3}$, although the $%
\Gamma _{15}$, mainly consisting in a displacement of Pb along $\left(
001\right) $,\cite{WR97} is stabilized in PbTiO$_{3}$ by the tendency of Pb
to shift in that direction in order to hybridize and form 4 short Pb-O
bonds; this is possible, since also the Ti ions shift off-centre from the
octahedra in the same direction, away from the O atoms with short Pb-O
bonds. In PbZrO$_{3}$ instead, where the ideal Zr-O bonds are longer than
the Ti-O ones, the rotational instability prevails in order to accommodate
the larger octahedra; this fact is usually expressed in terms of the
tolerance factor{\em \cite{HFS04}} $t=\left( r_{\text{A}}+r_{\text{O}%
}\right) /\sqrt{2}\left( r_{\text{B}}+r_{\text{O}}\right) $ which is 1 if
the ideal B-O distance matches the ideal A-O distance, and is $<1$ when the
octahedra are too large to fit the cell and tend to rotate to maintain their
size. The substitution of Pb with La should favor the rotational instability
because: {\it i}) it decreases the tolerance factor, due to a smaller ideal
La$^{+3}$-O bond length (2.76~\AA ) with respect to Pb$^{2+}$-O (2.89~\AA ),%
\cite{TEV96} and in fact XRD showed that the cell volume passed from 65~\AA $%
^{3}$ in PZT 20/80 to 63.4~\AA $^{3}$ in PLZT 22/20/80 (Fig. \ref{fig c/a});
{\it ii}) La$^{3+}$ does not have the ability of stabilizing the
ferroelectric tetragonal mode by a large electronic hybridization with O, as
Pb does.\cite{GCR02,TEV96}

We think that the best candidates for the non-polar motions we observe are
rotations of the octahedra related to the antiferrodistortive $R_{25}$ mode
described above. First-principle calculations of the phonon
dispersions showed that such a mode is unstable also in PbTiO$_{3}$,
although not enough to produce the ground state.\cite{GCW99} The dispersion
curves indicate a strong correlation of the cooperative rotations about $c$
within the $ab$ plane, due to the corner sharing of the octahedra, but
little correlation along $c$ (the same mode, with full correlation also
along $c$, is responsible for the cubic to tetragonal transformation in SrTiO%
$_{3}$\cite{Sco74}).\ Due to the lack of correlation between different
planes even in pure PbTiO$_{3}$ and to the strong disorder in the A\
sublattice in PLZT 22/20/80, the local condensation of such a mode over only
few octahedra should be possible, without producing superlattice peaks in
diffraction patterns. In addition, the small correlation length and
frustration from disorder would make it possible to switch between
differently rotated configurations. Switching between rotated and
non-rotated configurations would be accompanied by a change of the in-plane
shear strain (corresponding to $\Delta \lambda \neq 0$ in Eq. (\ref
{relstrength})) and therefore by anelastic relaxation, while the NMR
relaxation would be sensitive also to switching between correlated rotations
by opposite angles (which leave strain unchanged) and to the anharmonic
dynamics related to such an unstable mode.

Finally, the rotations of the octahedra are only weakly coupled to the
cation displacements in PLZT,\cite{CNI97} the major coupling perhaps coming
from the tendency of Pb to form four short Pb-O bonds.\cite{CGW98} In fact,
recent powder neutron diffraction experiments on rhombohedral PZT have been
interpreted assuming that, besides the established ferroelectric
displacements of Pb along the pseudocubic $\left\langle 111\right\rangle $
direction, additional $\left( 100\right) $ shifts of $\sim 0.2$~\AA\ are
induced by the tendency of Pb to form the fourfold coordination with the
surrounding O atoms, with the choice between $x$, $y$ or $z$ dictated by the
local tilt pattern of the octahedra.\cite{CGW98} Then, the weak coupling
between the ferroelectric cation displacements and the octahedra rotations
justify the fact that the latter are still fast at temperatures at which the
polar modes are completely frozen out; on the other hand, the weak low
temperature tail found also in the dielectric $\chi ^{\prime \prime }$ might
result from the supposed additional degree of freedom of Pb,\cite{CGW98}
which is coupled to the octahedra (besides the fact that disorder might
induce a polar component also to the rotations of the octahedra).

Further support to the hypothesis that rotations of the octahedra are
responsible for the low temperature NMR and anelastic relaxation comes from
the comparison of the anelastic spectra of PLZT 22/20/80 and PLZT 9/65/35,%
\cite{114} shown in Fig. \ref{fig tetrarhomb}. The latter composition is in
the rhombohedral region of the phase diagram, where the rotational
instabilities are documented, and in fact the low temperature component
alone determines the anelastic spectrum of PLZT 9/65/35, whereas it is
less important in PLZT 22/20/80.

\subsection{Antiparallel cation motion}

In principle, also correlated antiparallel motions of the A or B cations
might explain the discrepancies between the dielectric relaxation on one
side and the NMR and anelastic relaxation on the other side, since they
produce fluctuations both in shear strain and on the EFG at the La\ site,
but negligible dielectric response. Such modes are responsible for the
antiferroelectric phases in the Zr-rich end of the PLZT phase diagram,
but we do not see reasons why they should have importance in PLZT with 80\%
of Ti. In addition, such antiferroelectric modes involve displacements
of the Pb atoms;\cite{WK01} therefore, their condensation should be hindered
by the strong disorder in the A sublattice, and, if it occurred, it
would result in a polar mode, due to the charge variations among
different A sites (Pb$^{2+}$, La$^{3+}$ and vacancies). We
conclude that the extra contribution to the relaxation process
found in the anelastic and NMR experiments is rather due to modes
involving the O octahedra.

\section{CONCLUSION}

The combination of dielectric, NMR and anelastic spectroscopies in
the relaxor ferroelectric PLZT reveals that, besides the polar
fluctuations freezing at the relaxor transition, other
fluctuations of the atomic positions exist, that are faster and
weakly correlated with the polar ones. In fact, the imaginary
dielectric susceptibility $\chi ^{\prime \prime }\left( \omega
,T\right) $ presents the usual peak, with a frequency dispersion
signaling the freezing of the polar degrees of freedom, and a weak
tail at lower temperature, while the curves of the imaginary
elastic compliance $s^{\prime \prime }\left( \omega ,T\right) $
and especially of the $^{139}$La NMR relaxation rate $2W$ versus
temperature present a more intense component at lower
temperatures. This low temperature relaxation has been assigned to
non polar modes, which have been argued to correspond to rotations
of the octahedra; in particular, on the basis of the calculated
phonon dispersion curves in the PZT system,\cite{GCW99,FS01} it is
suggested that the condensation of the $R_{25}$ mode might lead to
clusters of rotated octahedra about the pseudocubic $c$ axis. The
correlation lengths for such correlated rotations should be very
short, due to the high concentration of La$^{3+}$ and Pb
vacancies, which hinder the otherwise predominant ferroelectric
tetragonal distortion. The small correlation length and the weak
coupling with the other ferroelectric modes would make it possible
for such clusters of octahedra to switch between differently
rotated configurations, giving rise to the observed low
temperature relaxation.

\section{Acknowledgments}

The authors thank A. Rigamonti for useful discussions and for his
advices. Thanks are also due to E.R. Mognaschi for his careful
check of the dielectric measurements and for useful discussions.

\section{Figure captions}

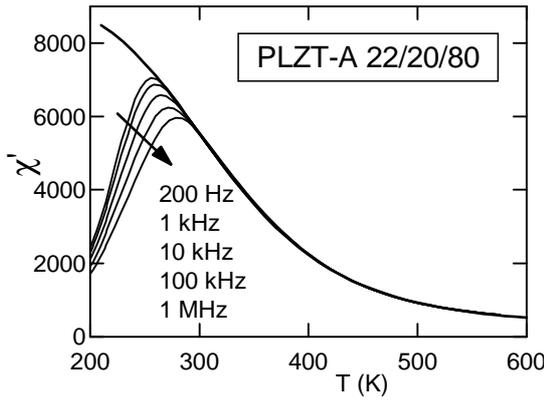
\begin{figure}[htb]
\caption{Real part of the dielectric susceptibility measured
at various frequencies. The thick line is $\chi \prime (0,T)$ adopted for
fitting $\chi(\omega,T)$.}
\label{fig chi(0,T)}
\end{figure}

\end{document}